# Universal behavior of the apparent fragility in ultraslow glass forming systems

Aleksandra Drozd-Rzoska

Institute of High Pressure Physics Polish Academy of Sciences,

ul. Sokołowska 29/37, 01-142 Warsaw, Poland

**Abstract**

Despite decades of studies on the grand problem of the glass transition the question of well-defined universal patterns, including the key problem of the previtreous behavior of the primary (structural) relaxation time, remains elusive. This report shows the universal previtreous behavior of the apparent fragility, i.e. the steepness index $m_P(T > T_g) = d\,log_{10}\,\tau(T)/d(T_g/T)$. It is evidenced that $m_P(T) = 1/(T - T^*)$, for $T \to T_g$ and $T^* = T_g - \Delta T^*$. Basing on this finding, the new 3-parameter dependence for portraying the previtreous behavior of the primary relaxation time has been derived:
$\tau(T) = C_\Omega \left((T - T^*)/T\right)^{-\Omega} \times \left[exp\left((T - T^*)/T\right)\right]^\Omega$. The universality of obtained relations is evidenced for glass formers belonging to low molecular weight liquids, polymers (melt and solid), plastic crystals, liquid crystals, resins and relaxors. They exhibit clear preferences either for the VFT or for the critical-like descriptions, if recalled already used modeling. The novel relation can obey even above the dynamic crossover temperature, with the power exponent $\Omega$ ranging between ~17 (liquid crystals) to ~ 57 (glycerol), what may indicate the impact of symmetry on the previtreous effect. Finally, the emerging similarity to the behavior in the isotropic phase of nematic liquid crystals is recalled.

The glass transition has remained the grand challenge of the solid state physics and material engineering since decades [1-9]. Its most attractive feature constitutes universal previtreous patterns for dynamic properties shared amongst the variety of glass forming systems, ranging from low molecular weight liquids to resins, melted polymers, solid polymers with segmental relaxation, liquid crystals, plastic crystals, colloids, …, and even granular systems [1-9]. The common key feature is the super-Arrhenius (SA) evolution of the primary relaxation time, viscosity, diffusion, …[8, 9]:

$$\tau(T) = \tau_0 \exp\left(\frac{E_a(T)}{RT}\right) \qquad (1a)$$

$$\eta(T) = \eta_0 \exp\left(\frac{E_a(T)}{RT}\right) \qquad (1b)$$

where $T > T_g$ and $T_g$ denotes the glass temperature; $\tau(T)$ and $\eta(T)$ are for the primary (structural) relaxation time and viscosity, respectively; $E_a(T)$ denotes the apparent (temperature dependent) activation energy being the key feature of the SA dynamics: for $E_a(T) = E_a = const$ the basic Arrhenius dependence is obtained.

Angell et al. [10, 11] proposed normalized plots $\log_{10}\eta(T)$ and $\log_{10}\tau(T)$ versus $T_g/T < 1$ enabling the common presentation of the SA previtreous behavior for different glass formers. To reach this target it was assumed that $\tau(T_g) = 100\,s$ and $\eta(T_g) = 10^{13}\,Poise$, what empirically coincides with the 'thermodynamic' estimation of $T_g$ from heat capacity studies with $10\,K/\min$ cooling rate. In the "Angell plot' the simple Arrhenius behavior appears as the straight line and the super-Arrhenius behavior via bend curves. As the universal metric categorizing the SA previtreous dynamics the slope $m = [d\log_{10}\eta(T)/d(T_g/T)]_{T=T_g}$ or equivalently $m = [d\log_{10}\tau(T)/d(T_g/T)]_{T=T_g}$, called fragility, was introduced [9, 10]. This report focuses on the evolution of $\tau(T)$.

The analysis of experimental data for the variety of glass formers, mainly low molecular liquids and polymers, led to the conclusion that they can be arranged within two categories [8-11]: (i) strong ($m < 30$) with the dynamics close to the basic Arrhenius pattern ($m \sim 16$) and (ii) fragile with the notably SA dynamics ($m > 30$). For the latter, values as large as $m \sim 200$ were reported. To describe such behavior beyond $T_g$ the steepness index, which can be considered as the apparent fragility metric, was introduced [8, 9]:

$$m_P = m_p(T) = \frac{d\,log_{10}\,\tau(T)}{d(T_g/T)} \qquad (2)$$

where $T > T_g$ and the subscript '$P$' indicates the isobaric character of $m(T)$ changes.

The simplicity matched with the premonition of a hidden universality caused that the 'Angell plot' has become the symbol of the mystery of the glass transition physics [1-9]. Parallel, studies related to the concept of fragility have become the central area of the glass transition physics [1, 6, 8, 9]. Nowadays, there is a set of model equations to portray $\tau(T)$ or $\eta(T)$ previtreous behavior, for instance they are the mode coupling theory (MCT [12]), Vogel-Fulcher-Tammann (VFT [8, 9, 13-15]), Avramov-Milchev (AM, [16]), Elmatad-Chandler-Garrahan (ECG. [17]), and Mauro-Yue-Ellison-Gupta-Allan (MYEGA, [18]), Kivelson-Tarjus-Zheng-Kivelon (KTZK, [19]), Schmidtke-Petzold-Kahlau-Hofmann-Rössler (SPKHR, [20]), … dependences. These relations are the reference and checkpoints for various theoretical frameworks proposed for explaining the glass transition phenomenon [1, 5, 6, 8, 9]. However, the general validity of these equations has been strongly questioned in the last decade. This is associated with their limited range of application, the validity only for selected systems or the lack of a relationship to other basic 'universal' properties of the glass transition [1, 8, 9, 21-26]. Particularly important seems to be the question of the singular temperature below $T_g$. This is the basic feature of the VFT equation [8, 9 13-15]:

$$\tau(T) = \tau_0 \exp\left(\frac{D}{T-T_0}\right) \qquad (3)$$

where $T > T_g$ and $T_0 < T_g$ is the VFT singular temperature; $D_T$ is the fragility strength coefficient linked to fragility: $m = m_{min} + \ln 10\, m_{min}/D_T$ and $m_{min} = log_{10}\tau(T_g) - log_{10}\tau_0$ [10]. For the prefactor most often the value $\tau_0 = 10^{-14}\,s$ is heuristically assumed [8, 9, 21].

The VFT singular temperature $T_0$ was linked to a phase transition 'hidden' below $T_g$ [1, 3, 8, 9] and to the ideal glass (Kauzmann) temperature $T_K$ [26, 27]. The latter is basically determined from the structural entropy and heat capacity studies [1, 8, 26]. The possible coincidence between the 'dynamic' ($T_0$) and 'thermodynamic' estimations of the hidden ideal glass transition has become one of the most important references in model studies on the glass transition [1-3, 8, 26]. However, the most extensive and challenging compilation of experimental data in ref. [28] showed that $0.8 < T_0/T_K < 2.2$, what means that $T_0 \approx T_K$ only for a limited group of glass formers. For several glass forming systems the residual analysis of experimental data showed the prevalence to AM, MYEGA or ECG descriptions [1, 29-31], avoiding the singularity below $T_g$. On the other hand the fundamental validity of AM or MYEGA relations was not supported by the apparent activation energy index analysis [23-25]. The concept of a possible 'hidden' continuous phase transition below $T_g$ lead also to the implementation of the relation originally developed for dynamic critical phenomena [32]:

$$\tau(T) = \tau_0(\xi(T))^z = \tau_0^C \left(\frac{T - T_C}{T_C}\right)^{-\phi = -zv} \qquad (4)$$

where $T_C < T_g$ is the critical temperature, the correlation length $\xi(T) = ((T-T_C)/T_C)^{-v}$ and $z$ is the dynamic (critical) exponent.

Such description proved its validity for spin-glass systems, where it can be clearly justified theoretically [33]. In this unique case Eq. (4) serves as the basic tool for estimating the glass transition temperature, because it is assumed that $T_C = T_g$ [33]. Regarding the 'critical' exponent values $\phi = 8 \div 13$ for spin glasses are reported [33]. Souletie and Bernard [34] tested the application of the 'spin-glass-type' description for molecular glass formers, but the presented analysis of experimental data exhibit notable discrepancies close to $T_g$. Murthy et al. [35 and refs. therein] applied the 'critical' Eq. (4) as the alternative to the VFT portrayal for a set glass forming plastic crystals. The decisive comparison of the formal validity of discussed descriptions was possible due to the linearized, derivative-based analysis [36-38], which proved the general preference of the critical-like Eq. (4) for glass forming liquid crystals and plastics crystals and the preference of VFT Eq. (3) for various low molecular weight systems. These tests yielded $\phi = 8.5 \div 11$ and $T_C \approx T_g - 10K$ for LC and $\phi = 8 \div 15$ and $T_C \approx T_g - 20K$ for ODICs [36-38].

Regarding the fragility concept, decades of studies resulted mainly in finding its correlations to many other characteristics of the glass transition. Nevertheless, there is still no unambiguous concepts making from the fragility the clear metric of universality linking microscopically different glass formers [1-9, 24, 25].

This report presents the discovery of the empirical, universal, previtreous anomaly related to the apparent fragility. It is used for the derivation of the new equation for portraying the previtreous behavior for $T \to T_g$, presumably universal for arbitrary glass former. All these reveals new features of the SA previtreous dynamics, including the link to collective dynamic heterogeneities.

**Methods**

Results presented are the consequence of the analysis of the primary (structural, alpha) relaxation time $\tau(T)$ basing on high resolution and 'high density' (the number of tested temperatures per decade) experimental broad band dielectric spectroscopy studies. They were carried out in a broad range of temperatures for selected glass formers ranging from low molecular weight liquids (glycerol) [36], polymers (polystyrene, polyvinyludene disulfide – PVDF) [39], resin (EPON 828: diglycidyl ether of bisphenol-A) [40], liquid crystals (5CB: pentylcyanobiphenyl) [41], 8*OCB (iso-octyloxycyanobiphenyl) [42] , disordered orientational crystals (ODIC, plastic crystals: the mixture of neopentylalcohol (NPA) and neopentylglycol (NPG)) [43] and the relaxor glass former in a hybrid system (30 % volume fraction dispersion of  $BaTiO_3$ microparticles in PVDF) [39]. Regarding liquid crystalline glass formers: 5CB exhibits the Isotropic – Nematic  (I-N) transition at $T_{I-N} \approx 304\,K$ and crystallizes at $T_{cryst.} \approx 290\,K$ [41]. Notwithstanding, the careful cleaning and degassing of the sample and the proper design of the measurement capacitor enabled the supercooling in the nematic phase down to the glass transition [41]. 8*OCB  can be cooled down to the glass temperature in the isotropic phase, at any cooling rate. It is the isomer  of octyloxycyanobiphenyl (8*OCB)  with the isotropic – nematic – smectic - crystal mesomorphism [42]. Glass temperatures of tested samples are given Table I and Table II in the Appendix. Results presented were obtained using the Novocontrol Broad Band Dielectric Spectroscopy (BDS) Analyzer, model 2015. The flat – parallel, gold – coated,  measurement capacitor with the bulk gap $d = 0.2\,mm$ ($U_{meas.} = 1V$) was placed in the Quattro temperature control unit. The primary relaxation time was determined from the peak frequency of primary relaxation process loss curves as $\tau = 1/2\pi f_{peak}$ [8].

**Results and Discussion**

Experimental $\tau(T)$ dependences for tested glass formers are presented in Figure 1. Notable is the manifestation of the isotropic - nematic (I-N) transition in the liquid crystalline 5CB at $T_{I-N} \approx 304K$ [41]. For the 'hybrid' relaxor the SA dynamics terminates at 244 K, where the crossover to the Arrhenius behavior on heating occurs [39]. The plastic crystal NPA-NPG melts at $T_m \approx 240K$ [43]. For each set the lowest measured temperature (the highest value of $1/T$) corresponds to the glass temperature value.

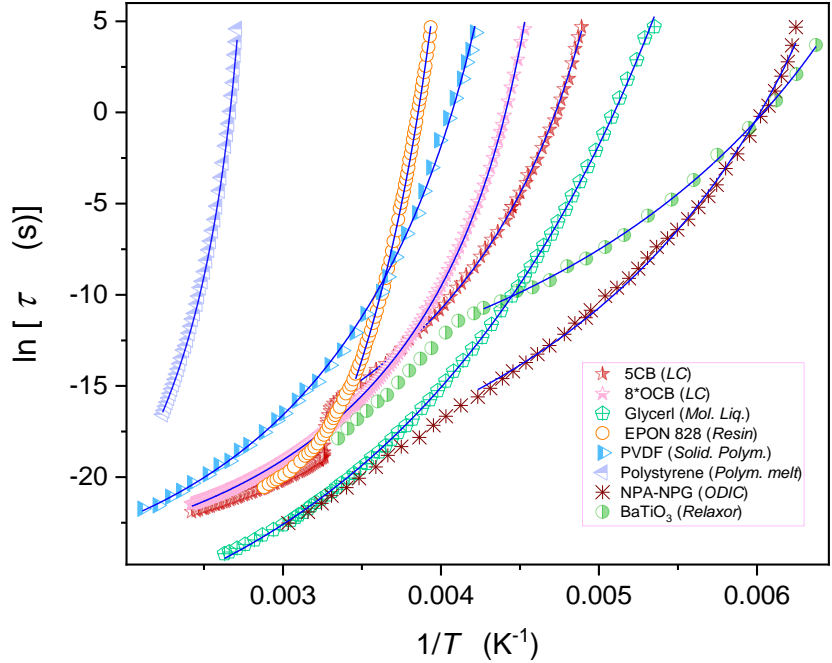

**Figure 1** Temperature dependences of the previtreous behavior $(T \to T_g)$ for glass forming 5CB, 8*OCB (LC), glycerol (low molecular weight liquid), EPON 828 (epoxy resin), polystyrene, PVDF (polymers), NPA-NPG (plastic crystal) and the 'hybrid' relaxor system: ~2 μm diameter BaTiO₃ ferroelectric particles (30% vol. f. in PVDF matrix. Solid curves are related to Eq. (7), with fitted parameters given in Table I (Appendix).

For these experimental data apparent fragilities (steepness indexes) were calculated using the basic definition (Eq. (2)), and the glass temperatures defined as $\tau(T_g) = 100\,s$. These results are presented in Figure 2. Tests of obtained $m_P(T)$ dependences led to the empirical finding:

$$\frac{1}{m_P(T)} = aT + b \quad (5a)$$

and then: 
$$m_P(T) = \frac{a^{-1}}{T - T^*} \quad (5b)$$

where $T^* = T_g - \Delta T^*$ and $\Delta T^*$ can be proposed as the measure of the 'discontinuity' of the glass transition; the singular temperature is determined via conditions $1/m_T(T^*) = 0$ or $T^* = -b/a$.

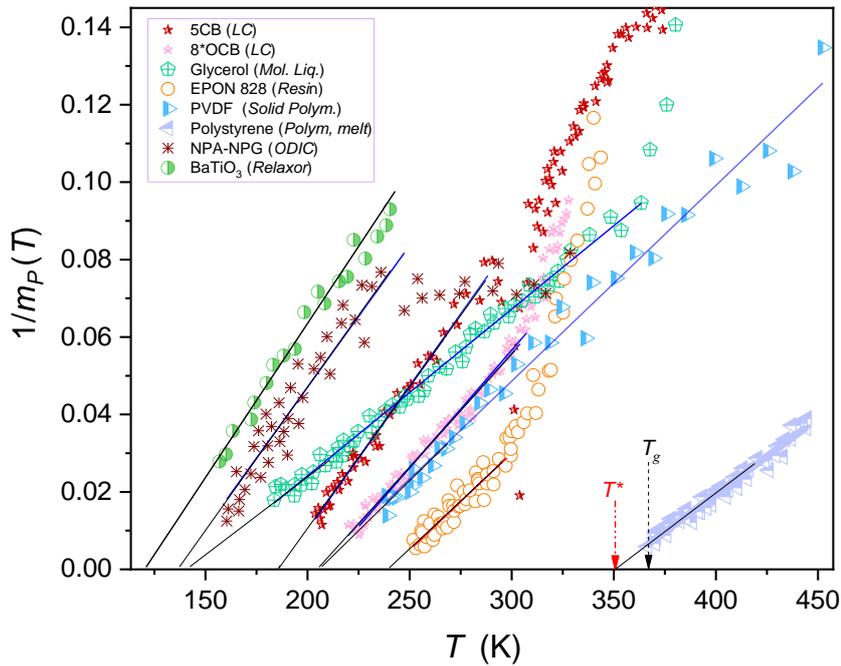

**Figure 2** Temperature dependences of reciprocals of steepness indexes (apparent fragilities) fitted by the linear dependence: $1/m_P(T) = a + bT$. Arrows indicate

locations of the glass temperature $T_g$ and the extrapolated 'pseudospinodal' singular temperature $T^*: [m_P(T^*)]^{-1} = 0$. Fitted parameters are collected in Table II (Appendix).

The hyperbolic dependence for the steepness index (Eq. (5)) occurs for all tested glass formers in the temperature range $T_g < T < T_{range}$, where the latter can reach even $T_g + 200 K$ for selected systems. It is noteworthy, that the transformation between the high temperature (ergodic) and the low temperature (non-ergodic) dynamical domains is gradual and smooth. Consequently, it seems that the simple and 'universal' behavior described by Eq. (5) emerges with first precursors of the non-egodicity. Linking the definition of the apparent fragility (Eq. (2)) and the empirical Eq. (5) one obtains:

$$m_P(T) = \frac{d \log_{10} \tau(T)}{d(T_g/T)} = \frac{a}{T - T^*} \tag{6}$$

The multiplication of both sides of Eq. (6) by $d(T_g/T)$ and the subsequent integration yields the new dependence for the primary relaxation time in the previtreous domain:

$$\tau(T) = C_\Omega \left(\frac{T-T^*}{T}\right)^{-\Omega} \left[exp\left(\frac{T-T^*}{T}\right)\right]^{\Omega} = C_\Omega \left(\frac{T-T^*}{T}\right)^{-\Omega} exp\left(\Omega \frac{T-T^*}{T}\right) \tag{7}$$

where $T > T_g$, $C_\Omega$, and $\Omega$ are constants; the extrapolated singular temperature $T^* = T_g - \Delta T^*$

The application of the new Eq. (7) for portraying the previtreous behavior of the primary relaxation times for various glass forming systems is shown in Figure 1, and fitted parameters are given in Table I, Appendix. Tested systems belong to different 'classes' of glass formers, as indicated above. Some of tested systems show the strong preference for the critical – like description (5CB, ODIC, ….) or the other for the VFT one (glycerol, EPON 828, PVDF. …). Notwithstanding, Eq. (7) offers the superior portrayal of the previtreous behavior for all tested glass formers even for $T > T_g + 200 K$. Notable is the fair agreement of $T^*$ values obtained for $\tau(T)$ and Eq. (7) nonlinear fitting and the linear regression fit of $1/m_P(T)$, as visible from the comparison of Table I and Table II (Appendix). This suggests that the number of fitted parameters in Eq. (7) can reduced to only 2, because $T^*$ can be easily estimated from the analysis of the reciprocal of the apparent fragility (Eq. (5a)).

On cooling towards the glass temperature two dynamical domains, associated with different forms of the SA dynamics, appears [8, 26, 44, 45]. They are separated by the dynamic crossover temperature $T_B$, for which the 'almost universal' time scale $\tau(T_B) = 10^{-7\pm1} s$ is suggested [46]. The existing broad experimental evidence shows that most often $T_B \sim T_g + 70K$ [8, 9, 36, 44, 45]. It is noteworthy, that the transformation between the high temperature (ergodic) and the low temperature (non-ergodic) dynamical domains is gradual and smooth [36, 44, 45]. Results presented shows the Eq. (7) can be applied even well above $T_B$: it seems that the simple and 'universal' behavior described by Eq. (7) emerges with first precursors of 'caged' relaxation of the low temperature dynamical domain.

**Conclusions**

Concluding, results presented show that the apparent fragility in the previtreous domain exhibits the universal behavior expressed via Eq. (5), what subsequently leads to the new relation for $\tau(T)$ portraying (Eq. (7)). It consists from two competing term: (i) the critical-like and (ii) the exponential one. Despite this fact it is associated with 3 fitted parameters, which can reduced to only 2, if the supporting by $[m_P(T)]^{-1}$ linear regression analysis is included. It is notable that both terms in Eq. (7) are associated with the same power exponent ranging between $\Omega \approx 17$ for liquid crystals, $\Omega \approx 27$ the plastic crystal and glassy relaxor, and $\Omega \approx 57$ for glycerol. Speculatively, one can consider this as the indication of the impact of the symmetry.

It is notable that Eq. (7) is associated with the 'relative' definition of the distance from the singular temperature: $(T-T^*)/T = (1-T^*/T)$. Such temperature scale appears in Eq. (7) in a natural way directly from the derivation Eq. (6). Worth recalling is the discussion regarding the 'absolute' $T-T_C$ or $(T-T_C)/T_C$ and 'relative' $(T-T_C)/T$ definitions of the distance from the singular (critical) temperature in the Physics of Critical Phenomena [47]. Although the 'relative' definition was indicated as fundamentally more optimal [47], the 'absolute' one is used in practice. This results from the fact that for critical phenomena the description via the leading power term $(T-T_C)^{-\alpha}$ is possible in the immediate vicinity of $T_C$, in practice for $(T-T_C)<1K$ and then there is no practical difference between both definitions. This is not the case of glass forming system where Eq. (7) can obey even up to $T_R \sim T^* + 120K$. It is notable that a similar 'relative' definition of the distance from the singular temperature appears for the 'thermodynamic' previtreous anomaly of the structural entropy $S_C(T) \propto (1-T_K/T)$ [1, 4, 5, 6, 8] or for its generalized form $S_C(T) \propto (1-T_K/T)^n$ [23-25]. In the opinion of the author the analysis of $\tau(T)$ or $\eta(T)$ experimental data in the ultraslowed / ultraviscous domain via Eq. (5) or Eq. (7) may offer the possibility of the simple and 'model free' estimation of the hypothetical ideal glass temperature. In this respects notable is the superior agreement between values of $T^*$ in Tables I and II (Appendix) and available 'thermodynamic' estimations and 'thermodynamic' estimations of $T_K$ for glycerol [8, 28]. In the opinion of the author the application of the VFT equation for determining $T_K$ ($\approx T_0$) in many cases is biased due to the limited-adequacy of the VFT portrayal for the given system.

Very recently, basing on the model numerical analysis of the vitrification process, Wang et al. [48] concluded: ... *We find that the time scale corresponds to the kinetic fragility of liquids. Moreover, it leads to scaling collapse of both the structural relaxation time and* *dynamic heterogeneity for all liquids studied, together with a characteristic temperature associated with the same dynamic heterogeneity….*'.

Then, one can consider the apparent fragility in the previtreous domain as the direct metric of dynamics for dynamical heterogeneities in the ultraslowed domain. Consequently, basing on results of this report one can indicate the notable similarity between the isotropic phase of rod-like nematic liquid crystals [41, 49] and the previtreous domain in glass forming systems [1-9]. For LC one can indicated [41, 49]: (i) the primary relaxation time for $T > T_{I-N}$ is described by equations (3) or (4) and the emergence of the SA description in this high temperature domain is associated with the impact of prenematic fluctuations-heterogeneities in the isotropic liquid ('fluidlike') surrounding, (ii) dynamics of prenematic fluctuations is described via $\tau_{fluct.} \propto 1/(T-T^*)$ and $T > T_{I-N} = T^* + \Delta T^*$ (see then Eq. (5)); (iii) the impact of prenematic fluctuations is very strong for methods related to the 4-point correlation functions (Kerr effect, nonlinear dielectric spectroscopy, ..) but for other methods such as density or dielectric constant can be almost negligible, due to the poor contrast factor between fluctuations and their surroundings; (iv) dynamic heterogeneities in supercooled liquids are related to the nanoscale time/space range and they are associated with no more than few tens of molecules, as shown experiments carried out ca. 10 K above $T_g$ i.e. $T - T^* \approx 40K$ (glycerol) [8, 9] but in the isotropic phase of 5CB one obtains the same parameters for prenematic fluctuations for $T - T^* \approx 40K$ [49, 50]. Notwithstanding, there is the basic difference the isotropic – nematic and the glass transition. The glass transition is 'stretched' in temperature and time whereas the I-N transition is the well-defined discontinuous phase transition. To comment these basic properties, one can indicate the clear difference between symmetries of prenematic fluctuations (heterogeneities) and their 'fluidlike' isotropic surrounding, associated with the similar 'sharp' difference between symmetries of neighboring isotropic liquid and nematic phases. This can be not the case of heterogeneities in a supercooled system above $T_g$: their structure can resemble the amorphous surrounding, but with larger solidity and eventually only weak symmetry distortions. Consequently the border between heterogeneities and their surrounding can be gradual and stretched what finally may lead to the 'stretched transition' to the amorphous solid state.

The progress in understanding and describing the anomalous previtreous increase of the primary (structural) relaxation time or viscosity is considered as the key to resolving the scientific challenge of the glass transition [1-9]. This report presents the new evidence (Eq. (7)) regarding both the evolution of the relaxation time and the apparent fragility.

**Acknowledgement**

This research was carried out due to the support of the National Centre for Science (Poland), project NCN OPUS ref. 2016/21/B/ST3/02203, head Aleksandra Drozd-Rzoska.

**APPENDIX**

Parameters resulted from fitting $\tau(T)$ experimental data via eqs. (7) and (8) for glass forming :

   (i)   liquid crystals 5CB (vitrification in the supercooled nematic phase and 8*OCB

   (ii)  glycerol (low molecular weight liquid)

   (iii) EPON 828 (epoxy resin)

   (iv)  Polystyrene (melt)

   (v)   PVDF (Solid. Segmental relaxation)

   (vi)  NPA-NPG (plastic crystal, ODIC)

Basic characterization for these compounds (full names, glass temperatures and fragilities) are given in the experimental section.

**Table I** Results of fitting of experimental data $\tau(T \to T_g)$, for the primary (structural) relaxation time via the new eq. (7), recalled also in the table.

| Fitted Relation | $ln\tau(T) = lnC_\Omega + \Omega \times \left(1 - \frac{T^*}{T}\right) - \Omega \times ln\left(1 - \frac{T^*}{T}\right)$ | | | | | | | |
|---|---|---|---|---|---|---|---|---|
| | Glass Forming Materials | | | | | | | |
| | | 5CB | 8*OCB | GLY-CEROL | EPON 828 | POLY-STYRENE | PVDF | NPA-NPG | RELA-XOR |
| Parameters | $lnC_\Omega[s]$ | -40.4 | -41.4 | -86.5 | -40.7 | -30.0 | -49.5 | -48.4 | -44.0 |
| | $T_g[K]$ | 205 | 220.7 | 189 | 253.9 | 369.6 | 237.4 | 160.3 | 155.0 |
| | $\Omega$ | 18.5 | 16.6 | 56.9 | 26.1 | 16.5 | 24.2 | 26.0 | 27.2 |
| | $\Delta T_{range}[K]$ | 80.7 | 194.1 | 179.5 | 41.1 | 77.4 | 217.0 | 283.7 | 80.7 |
| | $T^*[K]$ | 184.6 | 206.3 | 138.8 | 233.5 | 349.0 | 208.7 | 134.9 | 123.0 |
| | $\Delta T_{range} = (T_r - T_g)$, where $T_r$ denotes the terminal temperature of fitting | | | | | | | | |

**Table II** Results of the linear regression fit for the reciprocal of apparent fragile (steepness index) $[m_P(T)]$ in the ultraslowed previtreous domain close to $T_g$

| Fitted Relation | $m_p(T) = \dfrac{1}{aT+b}, \quad aT+b \neq 0$ | | | | | | | |
|---|---|---|---|---|---|---|---|---|
| | Glass Forming Materials | | | | | | | |
| | | 5CB | 8*OCB | GLY-CEROL | EPON 828 | POLY-STYRENE | PVDF | NPA-NPG | RELA-XOR |
| Parameters | $T_g[K]$ | 205.0 | 220.7 | 187.0 | 253.9 | 369.6 | 237.4 | 160.3 | 155.0 |
| | $\Delta T_{range}[K]$ | 83.2 | 86.1 | 176.4 | 42.3 | 75.4 | 215.6 | 284.1 | 284.1 |
| | $T^*[K]$ | 185.7 | 205.5 | 141.0 | 239.0 | 350.0 | 207.6 | 136.0 | 123.8 |
| | $\Delta T_{range} = (T_r - T_g)$, where $T_r$ denotes the terminal temperature of fitting | | | | | | | | |